\documentclass[a4paper,fleqn]{cas-sc}
\usepackage[square,numbers,sort&compress]{natbib}
\usepackage[capitalise]{cleveref}
\usepackage{mathrsfs} 
\usepackage[figurename=Fig.]{caption}
\usepackage{subfig}
\usepackage{todonotes}
\usepackage{float}

\ExplSyntaxOn              %%disable thumbnails
\keys_set:nn { stm / mktitle } { nologo }   %%disable thumbnails
\ExplSyntaxOff    %%disable thumbnails

\def\tsc#1{\csdef{#1}{\textsc{\lowercase{#1}}\xspace}}
\tsc{WGM}
\tsc{QE}
\tsc{EP}
\tsc{PMS}
\tsc{BEC}
\tsc{DE}
\begin{document}

%\listoftodos

\let\WriteBookmarks\relax
\def\floatpagepagefraction{1}
\def\textpagefraction{.001}

\shorttitle{UcVAE for Pattern Design in FP-Based Manufacturing}

\shortauthors{Q. Liu et~al.}

\title [mode = title]{Univariate Conditional Variational Autoencoder for Morphogenic Pattern Design in Frontal Polymerization-Based Manufacturing}                      

\cortext[corresponding]{Corresponding author}

\author[ncsa,bi,ae,ksu]{Qibang Liu}[orcid=0000-0001-7935-7907]
\ead{qibang@illinois.edu}
\credit{Writing – review \& editing, Writing – original
draft, Software, Methodology, Formal analysis, Conceptualization, Resources}

\author[mit]{Pengfei Cai}
\credit{Writing – review \& editing, Investigation, Conceptualization}

\author[ncsa, nyu]{Diab Abueidda}
\credit{Writing – review \& editing, Investigation, Resources}

\author[bi,ae]{Sagar Vyas}
\credit{Writing – review \& editing, Investigation, Conceptualization}

\author[ncsa,mse]{Seid Koric}
\credit{Writing – review \& editing, Investigation, Resources}

\author[mit]{Rafael Gomez-Bombarelli}
\credit{Writing – review \& editing, Supervision, Funding acquisition}

\author[bi,ae]{Philippe Geubelle}
\credit{Writing – review \& editing, Supervision, Methodology, Funding acquisition, Conceptualization}
\cormark[1]
\ead{geubelle@illinois.edu}

\affiliation[ncsa]{organization={National Center for Supercomputing Applications, 
University of Illinois Urbana–Champaign}, 
    city={Urbana},
    postcode={61801}, 
    state={IL},
    country={USA}}

\affiliation[bi]{organization={Beckman Institute for Advanced Science and Technology, 
    University of Illinois Urbana–Champaign}, 
        city={Urbana},
        postcode={61801}, 
        state={IL},
        country={USA}}

\affiliation[ae]{organization={Department of Aerospace Engineering, Grainger College of Engineering,
    University of Illinois Urbana–Champaign},  
        city={Urbana},
        postcode={61801}, 
        state={IL},
        country={USA}}

\affiliation[mse]{organization={Department of Mechanical Science and Engineering, Grainger College of Engineering,
University of Illinois Urbana–Champaign}, 
    city={Urbana},
    postcode={61801}, 
    state={IL},
    country={USA}}

\affiliation[nyu]{organization={Civil and Urban Engineering Department, 
New York University Abu Dhabi},
country={United Arab Emirates}}   

\affiliation[ksu]{organization={Department of Industrial and Manufacturing Systems Engineering, 
Kansas State University},
city={Manhattan},
postcode={66506}, 
state={KS},
country={USA}}    

\affiliation[mit]{organization={Department of Materials Science and Engineering, 
    Massachusetts Institute of Technology}, 
        city={Cambridge},
        postcode={02139}, 
        state={MA},
        country={USA}}

\begin{abstract}
    Under some initial and boundary conditions, the rapid reaction-thermal diffusion process taking place during frontal polymerization (FP) destabilizes the planar mode of front propagation,
    leading to spatially varying, complex hierarchical patterns in thermoset polymeric materials.
    Although modern reaction-diffusion models can predict the patterns resulting from unstable FP,
    the inverse design of patterns, which aims to retrieve process conditions that produce a desired pattern,
    remains an open challenge due to the non-unique and non-intuitive mapping between process conditions and manufactured patterns.
    In this work, we propose a probabilistic generative model named univariate conditional variational autoencoder (UcVAE)
    for the inverse design of hierarchical patterns in FP-based manufacturing.
    Unlike the cVAE, which encodes both the design space and the design target,
    the UcVAE encodes only the design space. In the encoder of the UcVAE, the number of training parameters is
    significantly reduced compared to the cVAE, resulting in a shorter training time while maintaining comparable performance.
    Given desired pattern images, the trained UcVAE can generate multiple 
    process condition solutions that produce high-fidelity hierarchical patterns.
\end{abstract}

\begin{graphicalabstract}
\includegraphics[width=\textwidth]{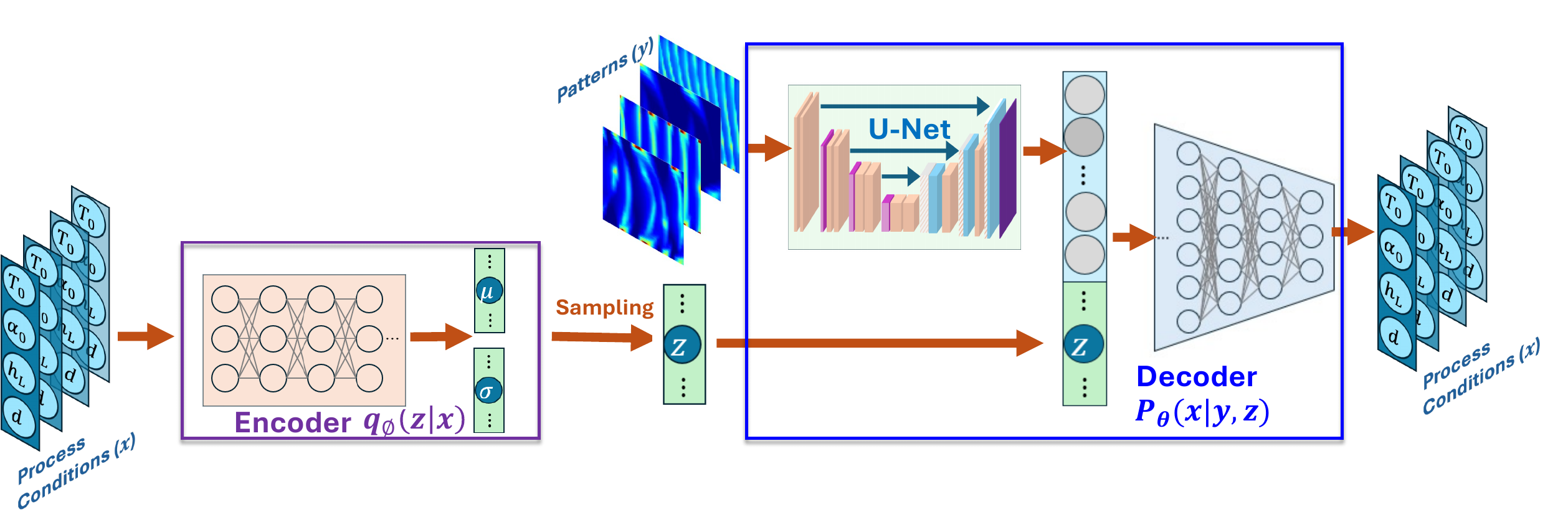}
\end{graphicalabstract}

\begin{highlights}
    \item A novel variational autoencoder (UcVAE) was proposed
    \item The UcVAE was applied to patterns design in FP-Based
    morphogenic manufacturing
    \item UcVAE saves 50\% training time comparing to cVAE for patterns design
\end{highlights}
% Keywords

\begin{keywords}
Frontal Polymerization
\sep Manufacturing \sep Inverse Design 
\sep Variational Autoencoder \sep Deep Learning 
\sep Deep Generative Model
\end{keywords}

\maketitle

\section{Introduction}\label{sec:intro}

In natural systems, structural complexity imparts organisms with unique characteristics 
that adorn biological materials and are critical to their survival.
These multiscale structural patterns originate from highly symmetrical initial states and develop during the growth and maturation of the organism \citep{melton1991pattern}.
Similarly, synthetic materials with hierarchical architectures that integrate the domains of hard and soft materials
along strong interfaces demonstrate superior properties compared to uniform architectures \citep{huang2019multiscale,rylski2022polymeric}.
The bio-inspired hierarchical patterns and structures in synthetic materials are usually created through multi-step 
manufacturing processes such as layer-by-layer assembly \citep{richardson2016innovation}, or additive manufacturing methods \citep{truby2016printing,saadi2022direct}, 
which require an a priori design and active intervention throughout the manufacturing process.
Other manufacturing processes involving autonomous routes based on synthetic coupled reaction-mass transport processes have been explored
\citep{campbell2004color,campbell2005one,karig2018stochastic,tan2018polyamide}, 
but these approaches are limited to solutions, gels, or thin membranes due to the prohibitively slow mass transport in solid media.

Recently, a novel manufacturing approach based on frontal polymerization (FP) was introduced, 
drawing parallels with morphogenic growth and development. This method enables the autonomous formation of architected pattern structures
within polymeric materials \citep{lloyd2021spontaneous,gao2021controllable,paul_controlled_2024,moore2024method}. 
FP is an out-of-autoclave, self-sustaining curing process that allows for
rapid and energy-efficient manufacturing of thermoset polymers \citep{pojman1996free,robertson2018rapid}.
Although stable front propagation in a homogeneous process configuration allows for the rapid production of thermoset polymeric materials \citep{mariani2001frontal,robertson2018rapid}, disturbances in the system, such as variations in initial and boundary conditions, can destabilize this planar mode. 
This destabilization can result in front propagation instabilities, leading to the formation of complex
hierarchical patterns within the material \citep{pojman1995spin,masere1999period,ilyashenko1998single}. For some materials, such as cyclooctadiene (COD), FP-induced instabilities result in a spatially varying thermal history, which leads to mechanical properties that can vary by multiple orders of magnitude \citep{lloyd2021spontaneous,paul_controlled_2024}.

Both stable \citep{goli2018frontal,goli_frontal_2020,vyas_frontal_2020,garg2021rapid,gao_frontal_2022,liu2024adaptive} and
unstable \citep{goli_instabilities_2020,gao2021manipulating,lloyd2021spontaneous,gao2021controllable,paul_controlled_2024} propagation 
of a polymerization front in a neat resin can be described with coupled, nonlinear thermo-chemical partial differential equations (PDEs). These equations are expressed in terms of degree of cure and temperature, along with the corresponding initial and boundary conditions.
Hierarchical patterns resulting from front instabilities can be obtained by solving the coupled PDEs with numerical methods, such as the finite element method (FEM), where the morphogenic pattern image is plotted based on the maximum temperature field
\citep{lloyd2021spontaneous,gao2021controllable,paul_controlled_2024}.

Although the effect of process conditions on hierarchical patterns has been studied \citep{paul_controlled_2024}, 
the inverse design of process conditions to achieve a desired pattern image remains an open challenge. 
Designing architectural structures for specific tasks is generally a complex problem, often involving a trial-and-error approach
or gradient-based optimization methods such as topology optimization \citep{andkjaer2010topology} or genetic algorithms \citep{huntington2014subwavelength}.
These methods are computationally expensive as they either use numerical methods such as FEM to solve PDEs iteratively and the performance of the search algorithms deteriorates rapidly as the design space expands. This challenge is particularly true for FP, where fine temporal and spatial discretizations are needed to capture the sharp gradients in temperature and degree of cure present near the advancing front. 

Unlike traditional optimization methods, machine learning (ML)-based data-driven approaches have shown great potential in solving inverse design
problems in various fields. Despite differing design targets and network architectures,
a common idea is to model the relationship between design parameters and targets as a bi-directional
mapping \citep{liu2018training,ma2018deep,gao2019bidirectional,goli2020chemnet}. 
However, such regression problems with one-to-one mappings are inconsistent with the physical intuition that different design parameters
can result in very similar design targets, which represents a one-to-many mapping.
Another common approach is to use deep-learning models as surrogate models to replace computationally expensive forward simulations for search by trial-and-error or gradient-based optimizations \citep{chugh2020surrogate,campbell2018advanced,hegde2019photonics,kudyshev2020machine,kollmann2020deep,abueidda2020topology}.
Once trained, the surrogate model can predict the design target based on the design parameters in milliseconds,
which is orders of magnitude faster than traditional numerical methods, thereby speeding up the inverse design trial-and-error process.
Furthermore, the gradient of the design target with respect to the design parameters can be quickly and easily obtained from the
surrogate model by automatic differentiation, accelerating the gradient-based optimization process. Although deep-learning surrogate
models can expedite the heuristics-based design process, they still require thousands of iterations to converge to the optimal solution.
More fatally, these models may have singularities in the design space, which could lead to optimization failures \citep{wiecha2021deep},
and the gradient-based optimization process may be trapped in local minima. Therefore, forward surrogate models for iterative optimization processes are not computationally efficient or robust enough for inverse design problems.

To address these challenges, numerous probabilistic generative models have been proposed for one-to-many inverse design problems,
such as conditional generative adversarial networks (cGAN) \citep{liu2018generative,so2019designing},
conditional variational autoencoders (cVAE) \citep{ma2019probabilistic,yonekura2021data},
and conditional diffusion models \citep{bastek2023inverse,weiss2023guided,vlassis2023denoising,igashov2024equivariant,park2024inverse,park2024nonlinear}.
These generative models can yield multiple solutions in the design space based on the design target
by learning the data distribution of the typically high-dimensional design space
and the usually low-dimensional design target space.
However, in the hierarchical pattern design problem in FP-based morphogenic manufacturing,
the design space, which describes the process conditions, is low-dimensional, whereas the design target, i.e., the pattern image, is high-dimensional.

To map one pattern image to many solutions of process conditions, we propose a novel probabilistic generative model called the univariate conditional variational autoencoder (UcVAE). Unlike the cVAE, which encodes both the design space and the design target, the UcVAE encodes only the design space. In the UcVAE encoder, the number of training parameters is significantly reduced compared to the cVAE,
resulting in a shorter training time while maintaining comparable performance.
Given pattern images from the test dataset generated by simulations of FEM or patterns extracted from real-world applications,
trained UcVAEs for FP-based morphogenic manufacturing design can generate multiple process condition solutions
that produce hierarchical patterns that closely match the target pattern images.

The paper is organized as follows: In \cref{sec:methods}, we describe the thermo-chemical reaction model and data generation process
using FEM, the architecture of the forward model that predicts the pattern based on the process conditions,
and both the cVAE and the proposed UcVAE models for inverse design.
In \cref{sec:results}, we present and discuss the results of the forward model and the inverse design models.
Finally, we summarize the work and future directions in \cref{sec:conclusion}.

\section{Methods}\label{sec:methods}

\subsection{ Thermo-chemical reaction model and data generation}\label{sec:fem}

\begin{figure}[hb]
    \centering
    \includegraphics[width=\textwidth]{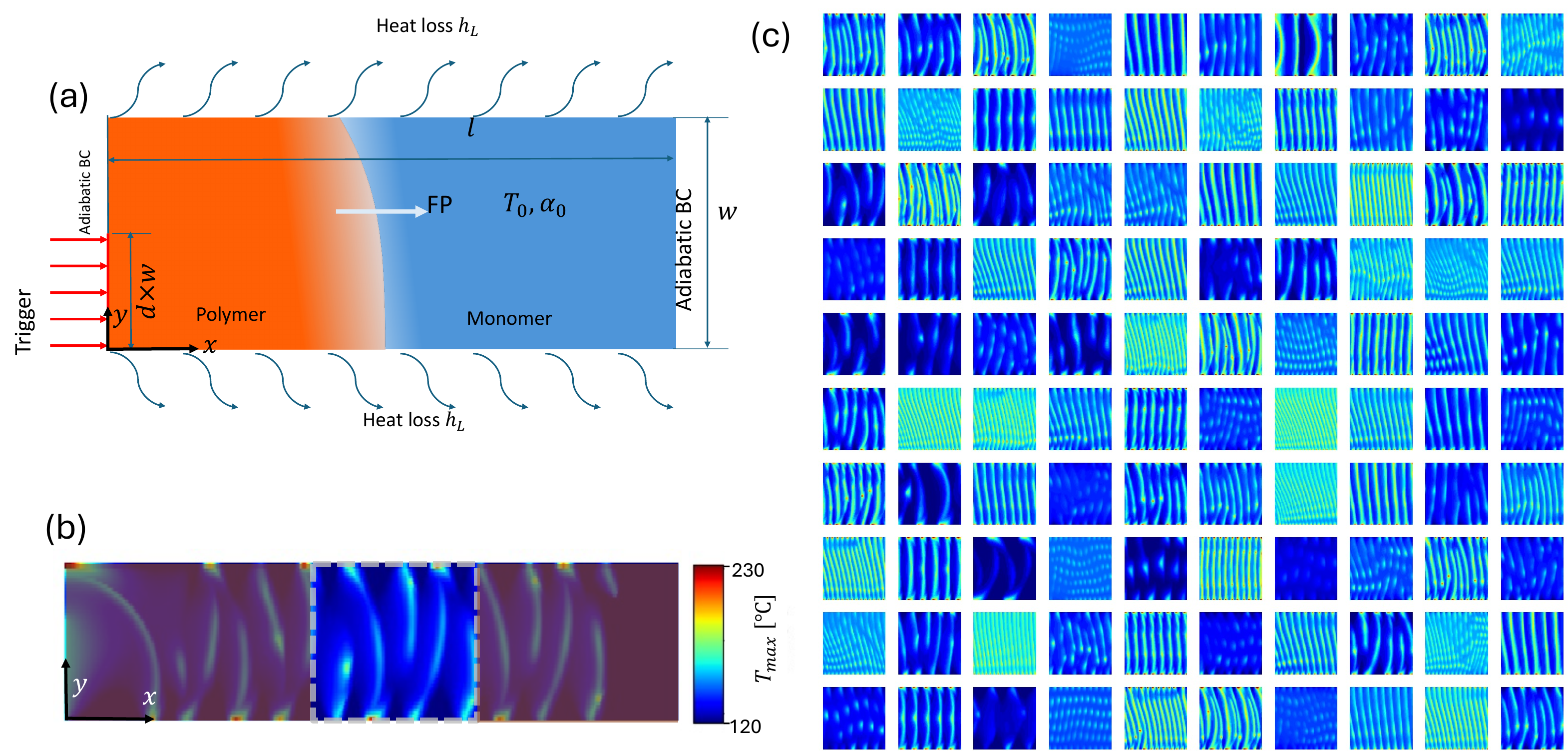}
    \caption{FEM simulation and data generation. (a) problem setup, 
    (b) window of interest, 
    (c) 100 typical pattern examples out of the 4752 total samples.}
    \label{fig:femdata}
\end{figure}

The frontal polymerization of cyclooctadiene (COD) can result in hierarchical patterns \citep{lloyd2021spontaneous,paul_controlled_2024}. 
The initiation and propagation of the polymerization front in COD can be described by the following coupled thermo-chemical
reaction-diffusion equations,
expressed in terms of the degree of cure $\alpha$ and temperature $T$:
\begin{equation}\label{eq:ge}
    \left\{\begin{array}{l}
        \kappa \nabla^2 T + \rho H_r \frac{\partial \alpha}{\partial t} = \rho C_p \frac{\partial T}{\partial t}, \\
        \frac{\partial \alpha}{\partial t} = A \exp \left(-\frac{E}{R T}\right) (1-\alpha)^n \alpha^m,
    \end{array}\right.
\end{equation}
where $\kappa$ is the thermal conductivity, $\rho$ is the density, $H_r$ represents the total enthalpy of reaction,
and $C_p$ denotes the specific heat capacity. The second equation in \cref{eq:ge} describes the cure kinetics of the FP reaction,
with $A$ representing the pre-exponential factor, $E$ the activation energy, and $R$ the universal gas constant.
We use the Prout-Tompkins cure kinetics model, $(1-\alpha)^n \alpha^m$, to describe the reaction order.
The cure kinetics parameters $A$, $E$, $n$, and $m$ are obtained from nonlinear fitting of heat flow measurements
from differential scanning calorimetry (DSC) tests \cite{lloyd2021spontaneous}.
The material properties and cure kinetics parameters for COD are listed in \cref{tab:const}.
\begin{table}[hbtp]
    \caption{ Cure kinetics and material properties of COD \citep{gao2021manipulating}.}
    \centering
    \begin{tabular}{cccc}
        \hline
        $\kappa\left(\frac{\mathrm{W}}{\mathrm{mK}}\right)$ 
        & $\rho\left(\frac{\mathrm{kg}}{\mathrm{m}^3}\right)$ 
        & $C_{p}\left(\frac{\mathrm{J}}{\mathrm{kg} K}\right)$ 
        & $A\left(\frac{1}{\mathrm{s}}\right)$ \\
        $0.133$ & 882  & 1838.5 & $2.13\times 10^{19}$\\
        \hline
        $E\left(\frac{\mathrm{kJ}}{\mathrm{mol}}\right)$ & $n$ & $m$ & $H_{\mathrm{r}}\left(\frac{\mathrm{J}}{\mathrm{g}}\right)$ \\
        $132$ & $2.514$  & $0.817$  & 220.0 \\
        \hline
    \end{tabular}
    \label{tab:const}
\end{table}

The FP initiation and propagation simulations presented in this manuscript pertain to the 2D problem shown
schematically in \cref{fig:femdata}(a). In this setup, a polymerization front is initiated by a thermal trigger
applied along the lower part of left edge with a length of $d \times w$ of a rectangular channel
with dimensions $l \times w$. Once initiated, the polymerization front propagates to the right. A convective boundary condition is applied along
the top and bottom of the reaction channel to capture heat loss to the surrounding.
The initial and boundary conditions used to complete the problem description are:
\begin{subequations}
    \begin{align}
         & T(x,y,0) = T_0, \quad \alpha(x,y,0) = \alpha_0, \quad 0 \leq x \leq l, \quad 0 \leq y \leq w \\
         & T(0,y,t) = T_{\mathrm{trig}}, \quad 0 \leq y \leq d \times w, \quad t \geq 0 \\
         & \frac{\partial T}{\partial x}(0,y,t) = 0, \quad d \times w < y \leq w, \quad t \geq 0 \\
         & \frac{\partial T}{\partial x}(l,y,t) = 0, \quad 0 < y \leq w, \quad t \geq 0 \\
         & \kappa \frac{\partial T}{\partial y}(x, y, t) = -h_L(T - T_0), \quad 0 \leq x \leq l, \quad t \geq 0, \quad y = 0 ~\text{or}~w.
    \end{align}
    \label{BCIC}
\end{subequations}
Here, $l = 20$ mm and $w = 5$ mm are the length and width of the channel, respectively,
while $T_{\mathrm{trig}} = 160~^\circ$C denotes the trigger temperature.
In this work, the process conditions are defined by the initial temperatures $T_0$, the initial degrees of cure $\alpha_0$, the thermal trigger length fraction $d$,
and the convective heat loss coefficient $h_L$.

\begin{figure}[b]
    \centering
    \includegraphics[width=\textwidth]{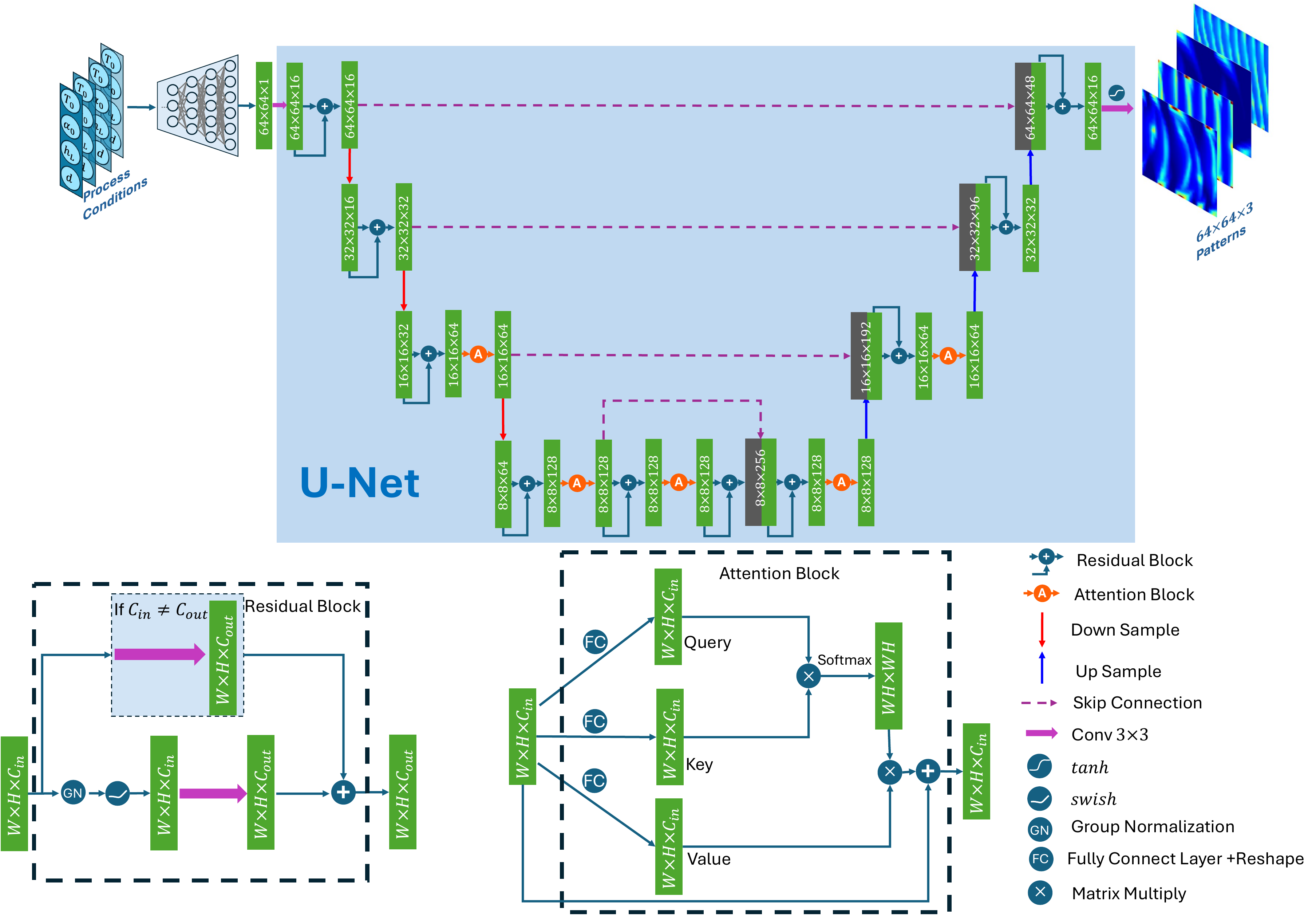}
    \caption{Forward model linking the four process conditions to the morphogenic pattern.}
    \label{fig:fwd_model}
\end{figure}

The coupled thermal-chemical reaction-diffusion equations in \cref{eq:ge} are solved using the finite element method (FEM) and the simulation is 
terminated when the front reaches 90\% of the channel length.
The maximum temperature $T_{\mathrm{max}}(x,y)=\max\limits_{t>0}{\{T(x,y,t)\}}$ is recorded at each point of the domain during the simulations.
Once the FP is completed, the hierarchical pattern is extracted based on the $T_{\mathrm{max}}$ field \citep{lloyd2021spontaneous,paul_controlled_2024} and plotted
using a color spectrum, as shown in \cref{fig:femdata}(b).
In this spectrum, dark blue represents the lower temperature bound of $120~^\circ$C and dark red represents the upper bound of $230~^\circ$C, with the number of intervals set to 30.
We then extract patterns (see \cref{fig:femdata}(b)) from the central part of the channel ($7.5~\text{mm} \leq x \leq 12.5~\text{mm}$) since the thermal trigger occurs at the beginning of the channel and the last 10\% of the channel is not cured. The pattern image in this region is then resized to $64 \times 64$ pixels
with three RGB channels.

Varying the process conditions, such as the initial temperature $T_0$, the initial degree of cure $\alpha_0$,
heat loss to the surrounding $h_L$, and the thermal trigger length $d$, results in different patterns.
We use the Sobol sequence with scrambling in the space of $T_0 \in [10,40]~^\circ$C, $\alpha_0 \in [0.01, 0.3]$,
$h_L \in [0,120] ~ W/m^2K$, and $d \in [0.3, 1.0]$ to sample the process conditions,
thereby generating 3538 patterns by solving the PDE with FEM. 
However, low values of $T_0$ and $d$ and high values of $\alpha_0$ and $h_L$ may lead to the quenching of front propagation, resulting in invalid patterns.
We filter out 1462 valid patterns, yielding a valid percentage of 41.3\%. 
We then estimate the mean value and the covariance matrix of the process conditions of the valid patterns to build a multivariate normal distribution.
Sampling process conditions from this distribution increases the valid percentage to 89.9\%. 
Finally, we obtain a total of 4752 samples, with \cref{fig:femdata}(c) showing 100 of them. 
Such data generation was performed using the FEM solver FeniCS \citep{alnaes2015fenics}, on {\fontfamily{qcr}\selectfont BEOCAT}, a Rocky Linux-based machine at Kansas State University. 
Approximately 100 simulations were run simultaneously with a total of \textasciitilde 800 CPUs, with each simulation taking around 6.3 hours with 8 CPUs.

\subsection{Forward model}

We first developed a forward model to predict the pattern image based on the process conditions 
to validate the results from the inverse design model.
To predict a high-dimensional pattern image ($64 \times 64 \times 3$)
based on low-dimensional process conditions with only four input parameters,
a fully connected neural network is first used to encode the process conditions.
The output is reshaped into a $64 \times 64 \times 1$ image, which is then fed into a U-Net to predict the pattern image.

The U-Net \citep{ronneberger2015u} is a convolutional neural network (CNN) architecture commonly used for image segmentation tasks.
To ensure stable training as the network depth increases,
residual blocks of convolutional layers \citep{he2016identity} were used to replace some of
the U-Net architecture's building blocks \citep{diakogiannis2020resunet}.
These residual blocks significantly mitigate the issue of vanishing and exploding gradients commonly
found in deep networks. Group normalization (GN) \citep{wu2018group} is used to normalize the feature maps,
and the swish activation function is used for nonlinearity in the residual blocks.
An attention mechanism \citep{vaswani2017attention} is incorporated at two low levels of the U-Net to adaptively learn attention-aware features.
The output of the U-Net is the pattern image with $64 \times 64$ pixels and three RGB channels.

The TensorFlow framework was used to implement the forward model,
and the Adam optimizer was used to minimize the mean squared error (MSE) loss function.
The schematic of the forward model, including the fully connected layers, U-Net, residual blocks, and attention blocks,
is shown in \cref{fig:fwd_model}. Further details of the implementation can be found in the 
\href{https://github.com/QibangLiu/UcVAE}{Github repository}.

\subsection{Inverse design}

\begin{figure}[b]
    \centering
    \includegraphics[width=\textwidth]{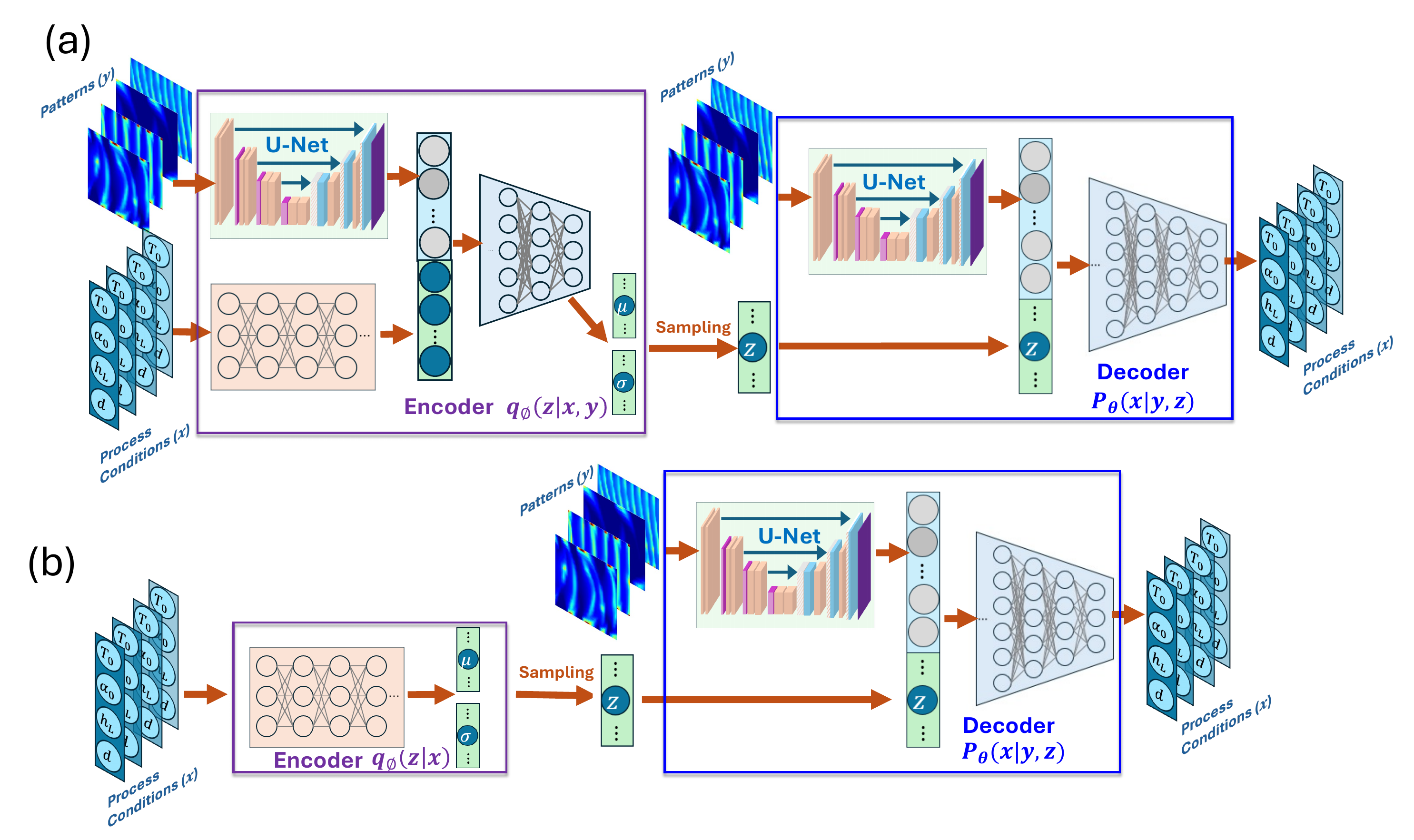}
    \caption{Neural network architecture for inverse design, where the U-Net architecture is the same as the one in the forward model.
    (a) Conditional variational autoencoder (cVAE) for inverse design,
    encoding both the design targets (patterns $y$) and the design parameters (process conditions $x$),
    originally proposed by \citet{sohn2015learning}. The encoder and decoder have 2,158,711 and 2,166,256 training parameters, respectively.
    (b) Univariate conditional variational autoencoder (UcVAE) proposed in this work for the inverse pattern design,
    encoding only the design parameters. The encoder and decoder have 644 and 2,166,256 training parameters, respectively.}
    \label{fig:inv_m}
\end{figure}
The major goal of this work is to develop a model for the inverse design of morphogenic manufacturing
patterns in FP due to thermo-chemical instabilities. Treating pattern design as a regression problem
with a one-to-one mapping contradicts physical intuition, as different process conditions can result in
very similar unstable responses and patterns, which is a one-to-many mapping.

To address this issue, we introduce latent variables as a probabilistic representation of pattern design
by incorporating a conditional variational autoencoder (cVAE) architecture, first proposed by \citet{sohn2015learning}.
As presented in \cref{fig:inv_m}(a), the cVAE encodes the process conditions ($x$) along with the
corresponding patterns ($y$) into a latent space ($z$).
The encoder defines a posterior distribution $q_{\phi}(z|x, y)$ (where $\phi$ denotes the parameters of the encoder),
which is used to sample latent variables $z$ from the latent space.
By imposing a specific prior distribution $p_{\theta}(z)$ on the latent space,
multiple designed process conditions can be reconstructed from the latent variables sampled within
the latent space, enabling these varied designs to produce similar target patterns.

The generative model, also named the decoder, defines the generative distribution of the process conditions
($x$) given the target patterns ($y$) and latent variables ($z$), $p_{\theta}(x|y, z)$ (where $\theta$ denotes
the parameters of the decoder), which is used for the inverse design in this work. Within the inverse design process,
a set of latent variables $z$ are sampled from the prior distribution $p_{\theta}(z)$,
and a set of process conditions are then generated by the decoder given the target pattern $y$
and the latent variables $z$.
These generated process conditions are fed into the forward model or FEM simulation
to obtain the designed pattern images, which should be similar to the given target pattern image.

The training objective of the cVAE model is to maximize the likelihood of the conditional probability $p_{\theta}(x|y)$,
given the training dataset, as described in \citep{sohn2015learning}:
\begin{equation}
    \max \{ \log p_\theta(x|y) \}=\max \{L_{vbl}(x, y; \theta, \phi)\},
\end{equation}
where $L_{vbl}(x, y; \theta, \phi)$ is the variational lower bound (VLB) and is defined as
\begin{equation}
    L_{vbl}(x, y; \theta, \phi)=\mathbb{E}_{q_\phi(z|x)}\left[\log p_\theta(x|y, z)\right]-\mathrm{KL}\left[q_\phi(z|x,y)|| p_\theta(z)\right].
\end{equation}
In this relation, $\mathrm{KL}$ denotes the Kullback-Leibler divergence, which describes the discrepancy between the two distributions. Training the cVAE and maximizing the VLB is equivalent to minimizing the negative VLB, which leads to the loss function of the cVAE model
\begin{equation}
    (\theta, \phi)=\arg \min _{\theta, \phi} \{-L_{vbl}(x, y; \theta, \phi)\}=\arg \min _{\theta, \phi}{L_{loss}},
\end{equation}
where the loss function $L_{loss}$ is defined as
\begin{equation}
    L_{loss}=||x-\Hat{x}||+\mathrm{KL}\left[q_\phi(z|x,y)|| p_\theta(z)\right].
\end{equation}
The term $||x-\Hat{x}||$ denotes the reconstruction loss, which
measures the difference between the input process condition and the generated process condition.

In this cVAE-based inverse design neural network architecture shown in \cref{fig:inv_m}(a),
we need to encode and decode the pattern images, whose dimension is as high as $64 \times 64 \times 3$.
U-Net architectures are a prevalent choice, and we use the same U-Net architecture as in the forward model.
The process condition is low-dimensional, with only four parameters,
so a simple fully connected neural network is used to encode the process conditions.
The output is concatenated with the output of the U-Net and then fed into another fully connected neural network,
whose outputs are the mean and standard deviation of the latent space distribution with a dimension of two.
In the decoder, the latent variables are sampled from the latent space distribution, concatenated with the output of the U-Net,
and fed into a fully connected neural network to output the predicted process conditions.

In this cVAE model, the encoder and decoder have 2,158,711 and 2,166,256 training parameters, respectively,
whereas the U-Net has 2.06 million training parameters and constitutes the major part of both the encoder and decoder. We use a large U-Net because a large model is needed to extract the features of the high-dimensional pattern image,
while the process conditions include only four parameters, so a small model is sufficient to encode the process conditions.
Thus, we propose a novel cVAE for inverse design, which only encodes the generative variables (i.e., process conditions $x$),
but not the conditions (i.e., pattern images $y$). We refer to this cVAE as univariate cVAE (UcVAE)
because only the generative variables are encoded, as shown in \cref{fig:inv_m}(b).

Since we do not encode the pattern images using the U-Net,
the encoder is a simple fully connected neural network with only 644 training parameters.
The encoder defines a proposed posterior distribution $q_{\phi}(z|x)$. 
The decoder of the UcVAE is the same as the cVAE, which defines a generative distribution $p_{\theta}(x|y, z)$,
and the true posterior is $p_{\theta}(z|x,y)$.
The KL divergence describes the discrepancy between the proposed and true posterior distributions as
\begin{equation}
    \mathrm{KL}\left[q_\phi(z|x)|| p_\theta(z|x,y)\right]=\mathbb{E}_{q_\phi(z|x)}\left[\log q_\phi(z|x)-\log {p_\theta(z|x, y)}\right].
\end{equation}

\noindent The optimization objective of the UcVAE model is also to maximize the likelihood of the
conditional probability $p_{\theta}(x|y)$, which is equivalent to minimizing the negative VLB.
The VLB for the UcVAE is derived as follows:
\begin{equation}\label{eq:ucve_vlb}
    \begin{aligned}
    & \log p_\theta(x|y) > \log p_\theta(x|y) - \mathrm{KL}\left[q_\phi(z|x) || p_\theta(z|x, y)\right] \\
    & = \mathbb{E}_{q_\phi(z|x)}\left[\log p_\theta(x|y)\right] - \mathbb{E}_{q_\phi(z|x)}\left[\log \frac{q_\phi(z|x)}{p_\theta(z|x, y)}\right] \\
    & = \mathbb{E}_{q_\phi(z|x)}\left[-\log \frac{q_\phi(z|x)}{p_\theta(x|y) p_\theta(z|x, y)}\right] \\
    & = \mathbb{E}_{q_\phi(z|x)}\left[-\log \frac{q_\phi(z|x)}{p_\theta(x|y)} \frac{p_\theta(x, y)}{p_\theta(z x y)}\right] \\
    & = \mathbb{E}_{q_\phi(z|x)}\left[-\log \frac{q_\phi(z|x)}{p_\theta(x|y)} \frac{p_\theta(y, z)}{p_\theta(z x y)} \frac{p_\theta(y)}{p_\theta(y, z)} \frac{p_\theta(x, y)}{p_\theta(y)}\right] \\
    & = \mathbb{E}_{q_\phi(z|x)}\left[-\log \frac{q_\phi(z|x)}{p_\theta(x|y)} \frac{1}{p_\theta(x|y, z)} \frac{1}{p_\theta(z|y)} p_\theta(x|y)\right] \\
    & = \mathbb{E}_{q_\phi(z|x)}\left[-\log \frac{q_\phi(z|x)}{p_\theta(z|y)} + \log p_\theta(x|y, z)\right] \\
    & = \mathbb{E}_{q_\phi(z|x)}\left[\log p_\theta(x|y, z)\right] - \mathrm{KL}\left[q_\phi(z|x) || p_\theta(z)\right] \\
    & = L_{vbl}(x, y; \theta, \phi).
    \end{aligned}
\end{equation}
The negative VLB leads to the loss function of the UcVAE model as
\begin{equation}
    L_{loss} = ||x - \hat{x}|| + \mathrm{KL}\left[q_\phi(z|x) || p_\theta(z)\right].
\end{equation}

Once the model is trained, the encoder is no longer needed.
The decoder can be transferred to a machine as a standalone model to infer multiple solutions ($x$) almost instantly
based on one target ($y$) but various latent variables ($z$), for specific generative tasks such as practical quick design optimization in industrial workflows.
In this work, we developed the UcVAE for FP-based morphogenic manufacturing design, but it can also be used for many other generative tasks.
For further details on the implementation of the cVAE and UcVAE in this work, please refer to our
\href{https://github.com/QibangLiu/UcVAE}{Github repository}.

\begin{figure}[!h]
    \centering
    \includegraphics[width=\textwidth]{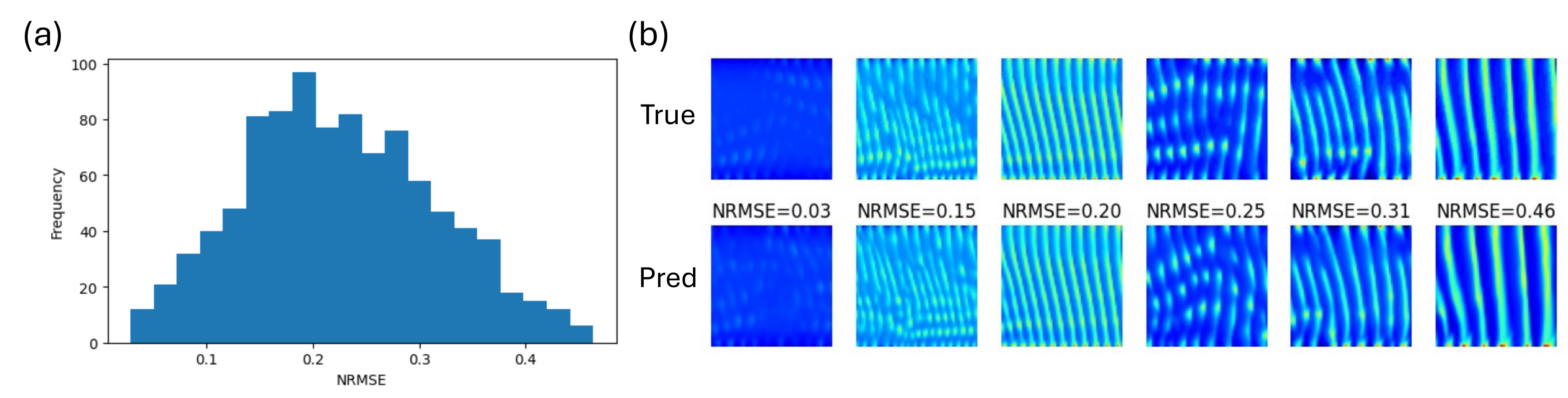}
    \caption{Forward model performance. (a) Normalized root mean squared error (NRMSE) of the test dataset.
    (b) Comparison between predicted and true patterns of the test dataset, arranged from best (left) to worst (right).}
    \label{fig:fwd_results}
\end{figure}

\section{Results and discussion}\label{sec:results}
In this section, we present the results of the forward model and the inverse design models. The total dataset generated using FEM simulation was shuffled and divided into 80\% training data and 20\% test data to train the
forward and inverse models.
Neural network training was performed on a single Nvidia A100 GPU on {\fontfamily{qcr}\selectfont DELTA},
a Red Hat Enterprise Linux-based machine at the National Center for Supercomputing Applications
at the University of Illinois Urbana-Champaign.
For model performance evaluation, the normalized root mean squared error (NRMSE) from the {\fontfamily{qcr}\selectfont skimage}
package is used:
\begin{equation}
    \mathrm{NRMSE} = \frac{||y^{\text{true}} - y^{\text{pred}}||_2}{y^{\text{true}}_{\text{max}} - y^{\text{true}}_{\text{min}}}.
\end{equation}

\subsection{Forward model performance}

The forward model is trained with 1600 epochs using the Adam optimizer with a learning rate of $5.0 \times 10^{-4}$
and a batch size of 128. Each epoch takes 2.2 seconds for training.
After training, it takes only 270 ms to predict the 951 test pattern images.
The distribution of NRMSE over all test samples is presented in \cref{fig:fwd_results}(a),
showing a maximum NRMSE of 0.46. The average value and standard deviation of the NRMSE distribution
for the test samples are 0.23 and 0.09, respectively. \cref{fig:fwd_results}(b)
displays a comparison between the true patterns and their corresponding predictions.
The patterns are evenly arranged from best (left image) to worst (right image).
As shown in that figure, even in the worst prediction with an NRMSE of 0.46, the forward model predicts the pattern image quite accurately.
This well-trained forward model is then used to evaluate the performance of the generative inverse model.

\subsection{Inverse design performance}
Two neural network architectures are developed for the inverse design of FP-based morphogenic manufacturing patterns.
One is based on the original cVAE architecture, which encodes both the process conditions and the pattern images,
with an encoder having 2,158,711 training parameters. The other is the proposed UcVAE architecture,
which only encodes the process conditions with an encoder having only 644 training parameters.
Both the cVAE and UcVAE are trained for 900 epochs using the Adam optimizer with a learning rate of $5.0 \times 10^{-4}$
and a batch size of 128.

\begin{figure}[htb]
    \centering
    \includegraphics[width=\textwidth]{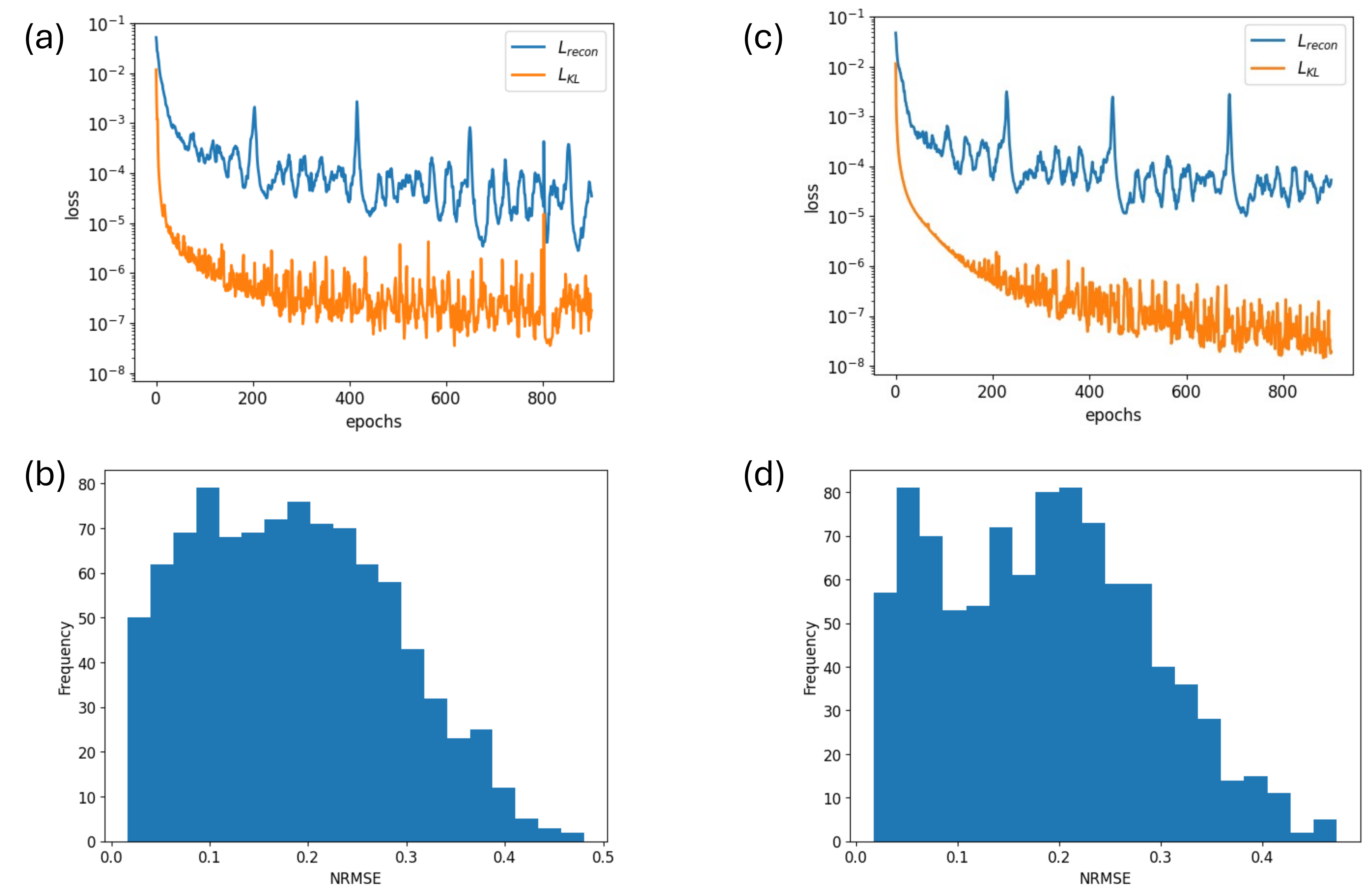}
    \caption{Model performance of inverse design.
    Loss values during the training process for cVAE (a) and UcVAE (c).
    NRMSE between target pattern image from the test dataset
    and corresponding designed pattern image obtained with cVAE (b) and UcVAE (d). In these figures,
    one random latent vector $z$ following a standard normal distribution is used to
    generate one process condition for each sample, and the process conditions are fed into the forward model
    to obtain the design images.}
    \label{fig:inv_results}
\end{figure}

\cref{fig:inv_results}(a) and (c) show the MSE loss values during the training process for cVAE and UcVAE, respectively.
As apparent in these figures, the reconstruction losses of both the UcVAE and cVAE are comparable, but the KL divergence of the UcVAE
is an order of magnitude lower than that of the cVAE.

Once the neural networks are trained, the designed process conditions are generated by
sampling one latent variable from the prior standard normal distribution for each design target
pattern sample in the test dataset. The designed process conditions are then fed into the forward model to obtain
the designed pattern images, which are compared with the corresponding design target pattern images.
The NRMSE distribution over the test targets is shown in \cref{fig:inv_results}(b) and (d) for cVAE and UcVAE, respectively.
As apparent in \cref{fig:inv_results}, the NRMSE distributions of cVAE and UcVAE are similar.
The average NRMSE value for cVAE is $0.1858$, which is slightly larger than that for UcVAE, $0.1849$.
The standard deviations for cVAE and UcVAE are also similar: $0.0995$ and $0.1014$, respectively.

\begin{figure}[!h]
    \centering
    \includegraphics[width=\textwidth]{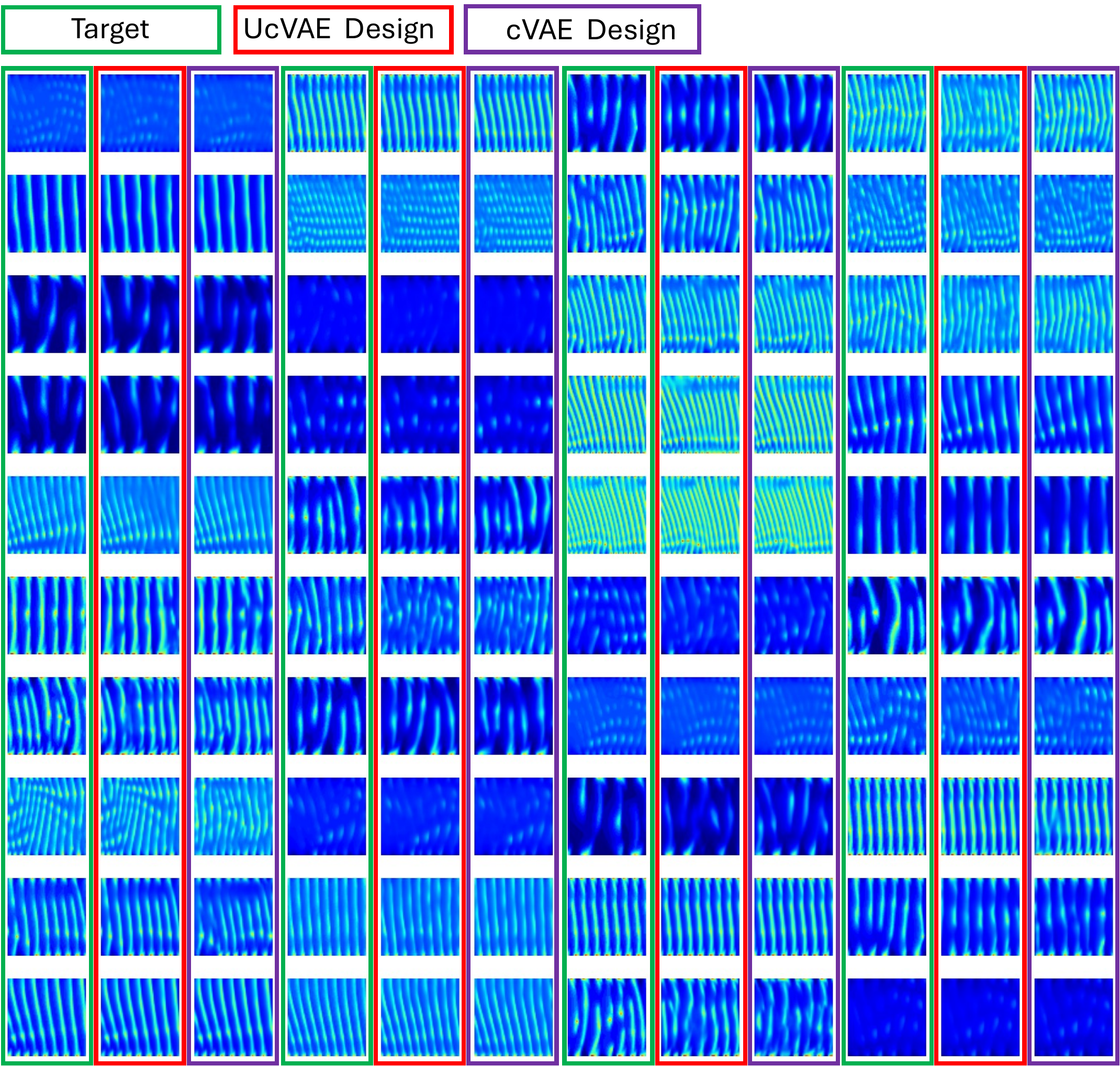}
    \caption{Forty random design examples obtained by UcVAE (red frames) and cVAE (purple frames). The targets (green frames) are random samples from the test dataset.
    One random latent vector $z$ following a standard normal distribution is used
    to generate one process condition for each sample,
    and the process conditions are fed into the forward model to obtain the design images.}
    \label{fig:inv_results2}
\end{figure}

We randomly selected 40 design targets from the test dataset and show in \cref{fig:inv_results2} the design results
obtained with cVAE and UcVAE. As apparent there, the two approaches yield predictions with patterns similar to the targets.
Both \cref{fig:inv_results} and \cref{fig:inv_results2} show that UcVAE and cVAE achieve similar design performance.
However, because UcVAE only encodes the process conditions into the latent space and has far fewer training parameters than cVAE,
training UcVAE takes only 1.8 seconds per epoch, which is 50\% less than the 3.6 seconds per epoch required for cVAE.

\subsection{One-to-many inverse design}
\begin{figure}[p]
    \centering
    \includegraphics[width=5.5in]{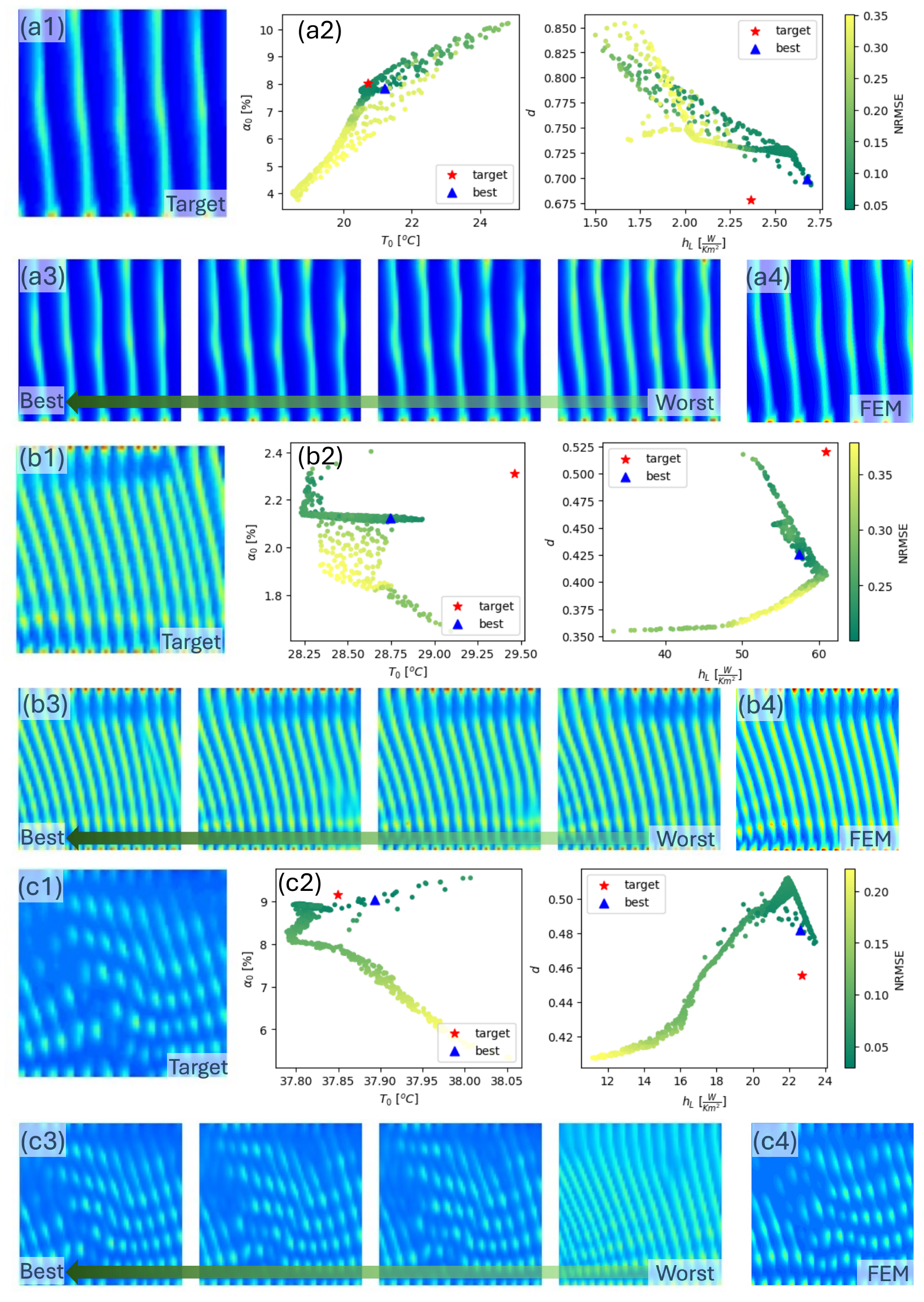}
    \caption{One-to-many inverse design of patterns using UcVAE.
    (a1), (b1), (c1) show random design targets from the test dataset.
    (a2), (b2), (c2) display 1024 designed process conditions generated
    by UcVAE with 1024 random latent vectors $z$ in Sobol space.
    (a3), (b3), (c3) present the designed pattern images predicted by the forward model with 
    the designed process conditions, arranged from best (left) to worst (right) in terms of NRMSE,
    where the best design process conditions are indicated by the blue triangle in (a2), (b2), (c2).
    (a4), (b4), (c4) show the designed pattern images obtained from FEM simulation with the best design process conditions.}
    \label{fig:one2many_test}
\end{figure}

\begin{figure}[p]
    \centering
    \includegraphics[width=6in]{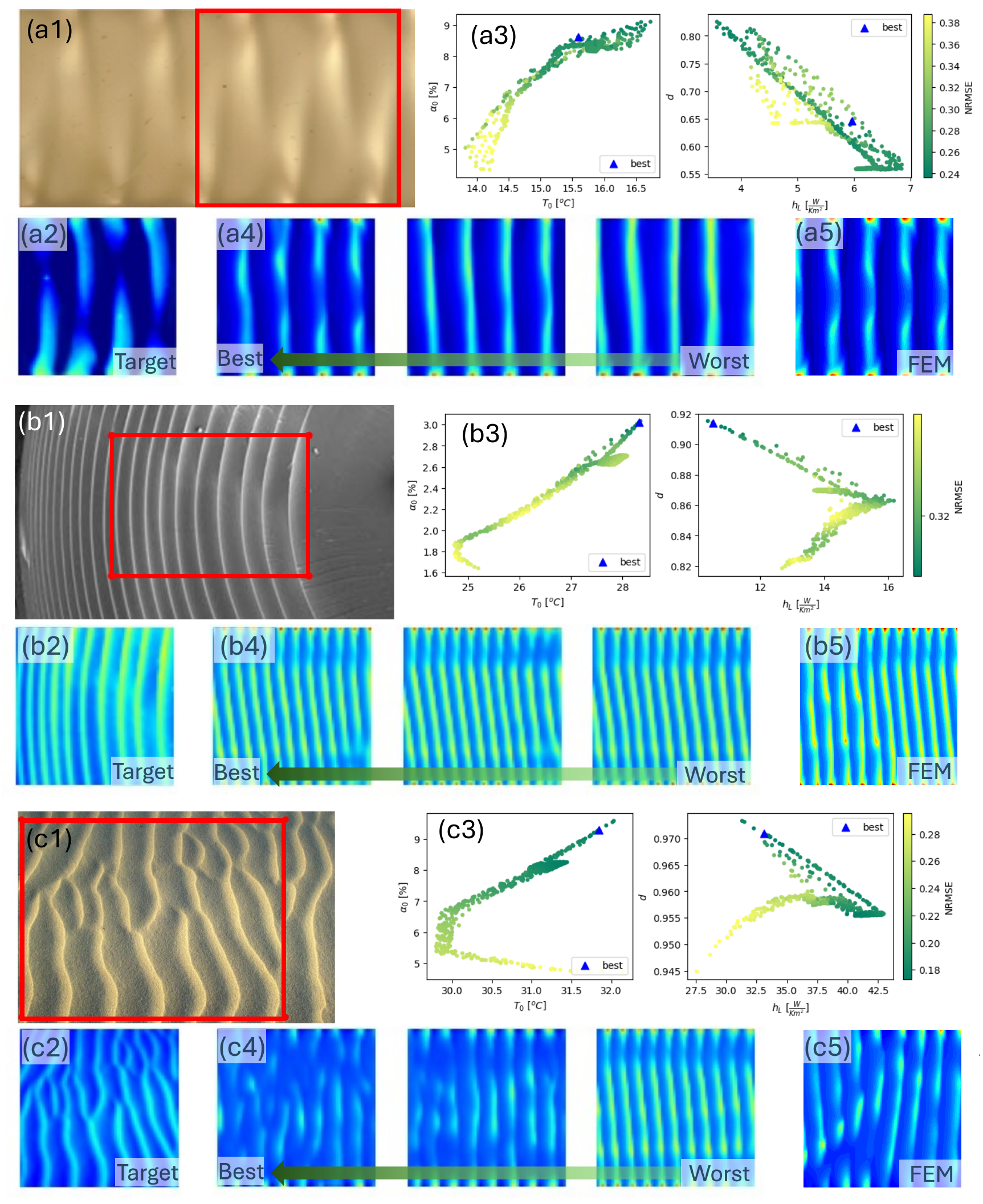}
    \caption{On-demand one-to-many inverse design of patterns using UcVAE.
    (a1), (b1), (c1) show real-world patterns from an experiment, an SEM image of a typical spicule,
    and an image of ripples of sand from nature, respectively.
    (a2), (b2), (c2) are the RGB design target patterns extracted from (a1), (b1), (c1), respectively.
    (a3), (b3), (c3) display 1024 designed process conditions generated by UcVAE with 1024 random
    latent vectors $z$ in Sobol space. (a4), (b4), (c4) present the designed pattern images predicted by the
    forward model with the designed process conditions, arranged from best (left) to worst (right) in terms of NRMSE,
    where the best design process conditions are indicated by the blue triangle in (a3), (b3), (c3).
    (a5), (b5), (c5) show the designed pattern images obtained from FEM simulation with the best design process conditions.}
    \label{fig:one2many_ondemand}
\end{figure}

The generative neural networks can be used for one-to-many solutions,
allowing multiple solutions to be generated for a given target.
In this section, we demonstrate the one-to-many inverse design of FP-based morphogenic manufacturing patterns using the UcVAE.
For a well-trained UcVAE, the encoder is no longer needed and only the decoder is used for inverse design.
For a given design target, changing the latent variables $z$ will result in different process conditions that
can produce designed images very close to the target image.
In the following inverse design tasks, the latent variables $z$ are not sampled
from a standard normal distribution but from a Sobol sequence with scrambling in the space of $[-64, 64]$
to achieve more diverse design solutions.

We randomly selected from the test dataset the 3 design targets shown in \cref{fig:one2many_test}(a1), (b1), and (c1).
For each design target, we sampled 1024 latent variables $z$ from the Sobol space. 
By feeding the design target and the 1024 latent variables into the decoder of the UcVAE,
we obtained 1024 designed process conditions for each design target, as shown in \cref{fig:one2many_test}(a2), (b2), and (c2).
The designed process conditions were then fed into the forward model to obtain the designed pattern images.
The NRMSE computed through comparison to the target is shown as a color map in \cref{fig:one2many_test}(a2), (b2), and (c2),
with the blue triangle representing the best design process conditions.
\cref{fig:one2many_test}(a3), (b3), and (c3) show the designed pattern images predicted
by the forward model with the designed process conditions, arranged from best (left) to worst (right) in terms of NRMSE.
As indicated in these figures, the designed pattern images associated with the three design targets and obtained from various designed process conditions
are very close to the target images.
\cref{fig:one2many_test}(a4), (b4), and (c4) show the designed pattern images obtained from FEM simulation
with the best design process conditions, which are also very close to the target images.
These one-to-many inverse design results demonstrate that
the UcVAE can generate multiple solutions for a given target.

While our developed UcVAE performs well on the test dataset for inverse retrieval,
real-world applications often demand on-the-fly inverse design of custom-defined patterns with high fidelity.
Therefore, instead of strictly achievable patterns such as those shown in \cref{fig:one2many_test},
a robust inverse design model should be able to generate possible designs from a rough sketch of predefined patterns.
Thus, instead of FEM simulation-generated patterns, we now use real-world patterns as design targets for on-demand inverse design. We adopt the patterns shown in \cref{fig:one2many_ondemand}.
\cref{fig:one2many_ondemand}(a1), (b1), and (c1). \cref{fig:one2many_ondemand}(a1) shows an experimentally observed FP-induced pattern described in \citep{paul_controlled_2024},
\cref{fig:one2many_ondemand}(b1) is a SEM image of a cross-section through a typical spicule in a laminated architecture taken from \citep{aizenberg2005skeleton},
and \cref{fig:one2many_ondemand}(c1) shows ripples of sand taken from nature.

These real-world pattern images are cropped along the red window shown in \cref{fig:one2many_ondemand}(a1), (b1), and (c1),
then converted to RGB images with color spectrum and resized to $64 \times 64$ pixels,
as shown in \cref{fig:one2many_ondemand}(a2), (b2), and (c2).
Similar to the inverse design of the samples from the test dataset shown in \cref{fig:one2many_test},
we sample 1024 latent variables $z$ from Sobol space for the design target from the real-world patterns,
and then feed the design target and the 1024 latent variables into the decoder of the UcVAE.
The designed process conditions are shown in \cref{fig:one2many_ondemand}(a3), (b3), and (c3),
where the blue triangle represents the best design process conditions.

Predicted design pattern images by the forward model with the designed process conditions
are shown in \cref{fig:one2many_ondemand}(a4), (b4), and (c4),
arranged from best (left) to worst (right) in terms of NRMSE.
\cref{fig:one2many_ondemand}(a5), (b5), and (c5) show the designed pattern images obtained
from FEM simulation with the best design process conditions. As apparent from \cref{fig:one2many_ondemand},
the designed pattern images from the real-world patterns are roughly close to the target images.
The difference in patterns is mainly due to the model being trained with FEM simulation-generated data,
whose distribution is quite different from the real-world patterns.

The three on-demand inverse design examples in \cref{fig:one2many_ondemand}
demonstrate the ability of the developed UcVAE to generate multiple solutions for
a given custom-defined target pattern, with the designed pattern images exhibiting rough fidelity to the target images.

\section{Conclusions and future work}\label{sec:conclusion}
In this work, we have proposed a novel probabilistic generative model named univariate conditional variational autoencoder (UcVAE) to conditionally generate
multiple solutions. Unlike the more classical conditional variational autoencoder (cVAE), which encodes both the solution space and the specific target,
the UcVAE encodes only the solution space. This significantly reduces the number of training parameters in the encoder,
resulting in a shorter training time while maintaining comparable performance.

We developed and trained a UcVAE-based neural network architecture incorporating U-Net, ResNet, and attention mechanisms
for the inverse design of hierarchical patterns in FP-based manufacturing.
The encoder of the UcVAE has only 644 training parameters, compared to the cVAE's 2,158,711 parameters.
Training the UcVAE takes only 1.8 seconds per epoch, which is 50\% less than the cVAE's 3.6 seconds per epoch,
while achieving similar design performance.
For a given design target from the test dataset generated by FEM simulation, the UcVAE can generate multiple solutions
that produce hierarchical patterns very close to the target images. Beyond strictly feasible patterns generated by FEM simulation,
three real-world patterns have also been used as the design target for on-demand inverse design.
In that more challenging case, the UcVAE also generated multiple solutions that produce hierarchical patterns with acceptable fidelity.

Beyond the inverse design of morphogenic manufacturing patterns in FP, the proposed UcVAE can be applied to many other generative tasks.
Future work include broadening the morphogenic manufacturing design space by incorporating more complex chemical compositions
and extending pattern designs from numerical simulations to other manufacturing processes.

\section*{Code availability}
The codes and data that support the findings of this study are available from the GitHub repository
at \url{https://github.com/QibangLiu/UcVAE}.

\printcredits

\section*{Conflict of interest}
The authors declare that they have no conflict of interest.

\section*{Acknowledgments}
This work was supported as part of the Regenerative Energy-Efficient Manufacturing of Thermoset Polymeric Materials (REMAT),
 an Energy Frontier Research Center funded by the U.S. Department of Energy,
Office of Science, Basic Energy Sciences under award DE-SC0023457. 

The authors would like to thank the National Center for Supercomputing Applications (NCSA) 
at the University of Illinois, and particularly its Research Computing Directorate, 
Industry Program, and Center for Artificial Intelligence Innovation (CAII) for support and hardware resources. 
This research is a part of the Delta research computing project, which is supported by the National Science Foundation (award OCI 2005572) 
and the State of Illinois, 
as well as the Illinois Computes program supported by the University of Illinois Urbana-Champaign and the University of Illinois System.

The data generation by FEM simulation for this project was performed on the Beocat Research Cluster at Kansas State University.

\bibliographystyle{elsarticle-num-names} 
%\bibliographystyle{cas-model2-names}
\bibliography{cas-refs}

\end{document}